\documentclass[amsmath,twocolumn,showpacs]{revtex4}
\usepackage{amssymb}
\usepackage{graphicx}
\usepackage{epsfig}
\begin{document}

\title{Effects of Quantum Tunneling in Metal Nano-gap on Surface-Enhanced Raman Scattering}
\author{Li Mao}
\affiliation{Institute of Physics, Chinese Academy of Sciences, Beijing
100190, China}
\author{Zhipeng Li}
\affiliation{Institute of Physics, Chinese Academy of Sciences, Beijing
100190, China}
\author{Biao Wu}
\email{bwu@aphy.iphy.ac.cn}
\affiliation{Institute of Physics, Chinese Academy of Sciences, Beijing
100190, China}
\author{Hongxing Xu}
\email{hxxu@aphy.iphy.ac.cn}
\affiliation{Institute of Physics, Chinese Academy of Sciences, Beijing
100190, China}
\date{\today}

\begin{abstract}
The quantum tunneling effects between two silver plates are studied
using the time dependent density functional theory. Our results show
that the tunneling  depends mainly on the separation and the
initial local field of the interstice between plates. The smaller
separation and larger local field, the easier the electrons tunnels
through the interstice.
Our numerical calculation shows that when the separation is smaller
than 0.6 nm the quantum tunneling dramatically reduces the enhancing
ability of interstice between nanoparticles.

\end{abstract}
\maketitle

Metal nano-gaps offering strong surface plasmon couplings have very 
rich physical properties. The related studies have been very hot topics in the 
field of plamonics, e.g., single molecule surface-enhanced Raman 
spectroscopy \cite{Xu1,Mich},  optical nano-antennas \cite{Muhlschlegel2005}, 
high-harmonic generation \cite{Kim2008}.
The electromagnetic (EM) enhancement near the metal surface, 
which is caused by the resonant excitation of surface plasmon
\cite{Raether}, is the dominating reason for the surface-enhanced
Raman scattering (SERS) \cite{Mosk,Xu1}. Huge SERS  with 
single molecule sensitivity can be obtained when molecules are located 
in the nano-gap between two metallic nano-structures \cite{Xu1,Mich,Xu2000,Talley,HWei}. 
A lot of efforts have been made to
seek extreme sensitive SERS substrates \cite{Haes1,Haes2,KZhao}. 

Theoretically,  people have used many methods based
on the classical electrodynamics \cite{Xu2007,Flatau, Futamata} to 
estimate the  SERS enhancement.
These classical results indicate that the smaller the nano-gap, 
the higher the enhancement. 
However, as the separation decreases to 1 nm, the displacive
current would partly become  electron tunneling current which can
reduce the EM enhancement substantially \cite{Otto}. A recent
experiment on the four-wave mixing at coupled gold
nanoparticles clearly demonstrated that the quantum tunneling (QT)
effect becomes significant for the distance smaller than 0.2 nm
\cite{Danck}, and a recent study of the plasmon resonance of a nanoparticle dimer gave 
quantum description of such a phenomenon\cite{Nordlander}. It is well known that the EM 
enhancement is the main contribution to SERS. Its enhancement factor is proportional to 
the fourth power of the local field enhancement, i.e. $M^4$, where $M$=$|E_{loc}|/|E_0|$ 
 with $E_{loc}$ and $E_0$ being the local enhanced electric 
field and the incident electric field, respectively. Therefore, even for 
small  QT effects on  $M$, after a fourth power, the influence to SERS 
could be huge. In this Letter we investigate the effects of QT on
SERS with the time dependent density functional theory \cite{RG}.
Our studies are able to quantify these effects and point out at
exactly what conditions the QT has to be taken into account.

\begin{figure}[t]
  \includegraphics[width=5cm]{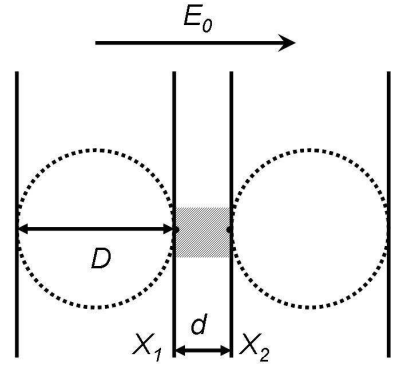}
  \caption{Schematic drawing of the ``hot spot" between two silver nano-spheres.
 As  the ``hot spot" (shaded area) is small,  its local field is almost identical to the
 one computed by replacing the sphere with a plate. $E_0$ is the incident laser field. }
\label{1}
\end{figure}
As the ``hot spot", where the SERS is strongest,  is localized in a
very small volume in the interstice between particles, it is
convenient to investigate the QT effect between two closely placed
plates instead of two nano-particles. As shown in Fig. 1, in the
vicinity of the ``hot spot" (shaded area),  two plates are not much
different from two nano-spheres. Besides we use two approximations
for our numerical calculations: (1) In the generalized Mie theory, the 
electric current inside nano-sphere is set to be zero \cite{Xu2007}, 
so we can regard the silver plates as
equipotential bodies at all time in our calculation; (2) The laser field is 
treated as a static electric field, and the QT effect in an oscillating 
field can be described by the results of static field in one period of laser. 
With these simplifications, when the separation $d$ is not very small, the
electric tunneling effect can be studied by the method developed by
Simmons \cite{simmons}, which regards electrons are tunneling
through a voltage barrier.  We  find that
Simmons' method is not proper when the distance $d<1$ nm. For example,  
at $d=0.6$ nm, the mean barrier
height becomes negative at low voltage limit, indicating the failure
of this method.

In this work we adopt a more sophisticated method, the time
dependent density functional theory (TDDFT) \cite{RG} with the
jellium model, where the ionic lattice is treated as a uniform
positive charge background.  In this method, we solve
self-consistently a set of time dependent Schr\"{o}dinger equations,
\begin{equation}
\label{main}
i\frac{\partial}{\partial t}\psi_{k}=\Big[-\frac{1}{2}\frac{\partial^{2}}{\partial
 x^{2}}+V_{eff}(x,t)+V_{ext}(x,t)\Big]\psi_{k} \,,
 \end{equation}
where we have used the atomic units and $\psi_k$ denotes a quantum state inside the Fermi
surface of the silver plate.  $V_{ext}(x,t)$ is the external potential coming from the
laser field and its induced field. $V_{eff}(x,t)$ is the effective potential felt by
an electron through Coulomb interaction and correlation and
exchange; it depends on the electron density. In our approach, 
we use Crank-Nicholson method \cite{cranknicholson}
to update the wave function.  
To quantify the QT effects on the SERS,
we monitor time evolution of the potential difference $\delta V$ between
the two silver plates. We compute $\delta V$ with the formula
\begin{eqnarray}
\delta V(t)=V_{eff}
(\frac{x_r^l+x_r^r}{2},t)-V_{eff}(\frac{x_l^l+x_l^r}{2},t)\,,
\end{eqnarray}
where $x_{r,l}^l$ and $x_{r,l}^r$ are coordinates of the left and right surfaces
of the right (left) plate.

Let us now turn on the laser field.
The electrons inside each silver plate will start moving instantly to counter-balance
the applied electric field so that the total electric field inside each plate
is zero. At the same time, an enhanced field is induced in the ``hot spot".
Afterwards, the electrons will start to tunnel between the two silver 
plates under the following external potential
\begin{equation}
 V_{ext}(x,t)=\left\{ \begin{array}{ll}
  E_0x & x<X_1\,,  \\
  E_{loc}(x-X_1)+E_0X_1 & X_1 \leqslant x  \leqslant X_2 \,,\\
  E_0(x-d)+E_{loc}d & x>X_2\,.
                   \end{array} \right. \\
\end{equation}

It is clear from the above analysis that the initial electron state
for the Schr\"odinger equations in Eq.(\ref{main}) is the state
where the electrons have moved to counter-balance the incident laser
field. To obtain this initial state, we compute with the method
developed by Schulte \cite{FK} the ground state of the metallic
plate under the following external potential
\begin{equation}
 V^0(x)=\left\{ \begin{array}{ll}
            0 & x<x^l \\ E_0(x-x^l) & x^l<x<x^r \,,\\ E_0D & x>x^r
                   \end{array} \right. \\
\end{equation}
where $x^l$, $x^r$ are the left and right surfaces of the plate.

\begin{figure}[h]
  \includegraphics[width=8cm]{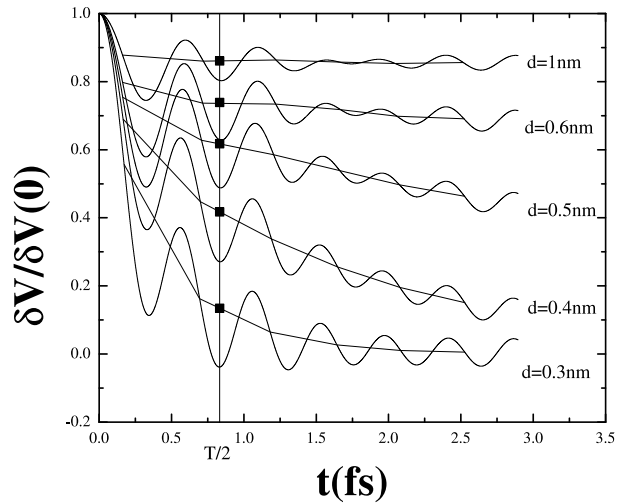}
  \caption{Time evolutions of the potential difference $\delta V$ between
the silver plates for different separations $d=0.3,\ 0.4,\ 0.5,\ 0.6$ and 1 nm.
 $D=6$ nm;  $E_0=2.74\times10^5$ V/m;  $M=1000$.
  Dashed lines are for the averaged $\delta \widetilde{V}$. }
  \label{2}
\end{figure}

Figure \ref{2} shows the
calculated time evolution of the potential difference $\delta V$
between the two plates separated by $d=0.3$ $\sim$ $1$ nm. 
The strength of the incident electric field is $E_0=2.74\times10^5$ V/m, 
corresponding to a laser with power $P=100~\mu W$ and focal spot 
$\sim 1~\mu$m. In most SERS experiments, even for single molecule 
detection, a much smaller $P\sim 1~\mu$W is used \cite{Xu1,Mich}. 
The diameter of nano-particle is $D=6$ nm.  Note that 
we have calculated for three 
different diameters $D=4,$ $5,$ $6$ nm and the results  are almost identical . 
This indicates that the physical process in 
the ``hot spot" is not sensitive to geometric features that are far away,
further justifying our replacement of the spheres with the plates.

We see in Fig. \ref{2} that $\delta V$ decays while oscillating with 
a frequency close to the bulk plasma frequency. The decay gets
severe as the separation becomes smaller.  This kind of decay can be intuitively
understood by viewing the system as a bad capacitor that leaks current. 


\begin{figure}[h]
  \includegraphics[width=8cm]{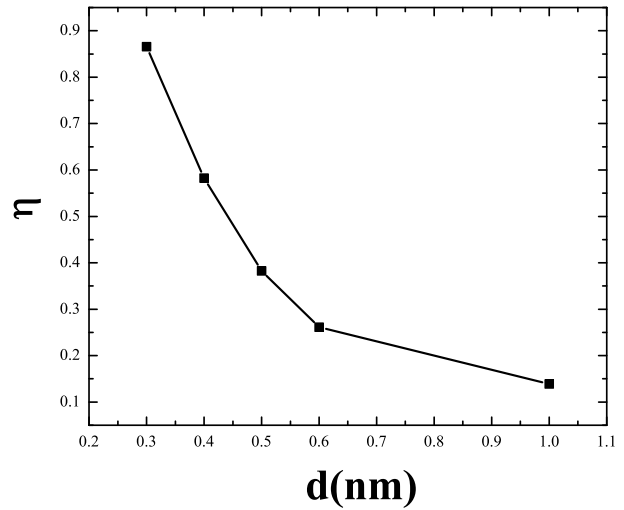}
  \caption{The decay rate $\eta$ of $\delta V$ as a function of
 the separation $d$. $D=6$ nm; $E_0=2.74\times10^5$ V/m;
 $M=1000$; $\lambda=500$ nm.}
\label{3}
\end{figure}


To measure the decay, or the QT suppression 
of the enhanced field, we introduce a decay rate defined by
\begin{eqnarray}
\eta=1-\frac{\delta \widetilde{V}(\frac{T}{2})}{\delta V(0)}
\end{eqnarray}
where $T$ is the typical optical period of the incident laser, e.g.,
$T=1.67$ fs for a laser wavelength $\lambda=500$ nm. Note that
$\delta \widetilde{V}(\frac{T}{2})$ is not the value of $\delta V$ 
but the averaged value of $\delta V$ over one
oscillation period at $t=T/2$. 
As the distance $d$ decreases, the potential
difference decays with time dramatically. At $d=0.3$ nm, the local
field is reduced by $\sim 86.6\%$ after half optical period
($\lambda=500$ nm), that is, the SERS enhancement is $3.1\times10^3$
times  ($(1-\eta)^{-4}$) smaller than the one obtained from classical theory. By 
contrast, at $d=1$ nm, the reduction of the local field
enhancement by the QT is only 14\%, corresponding to one time
decrease of SERS enhancement factor. This means that  the
enhancement can be sustained if the separation is larger than 1 nm.
The decay rates $\eta$ for these different separations
are computed and plotted in Fig.\ref{3},  where we see $\eta$
decreases exponentially as $d$ increases. Specifically,  when the
separation is smaller than 0.6 nm, the QT can reduce the local field
significantly. 


\begin{figure}[h]
  \includegraphics[width=8cm]{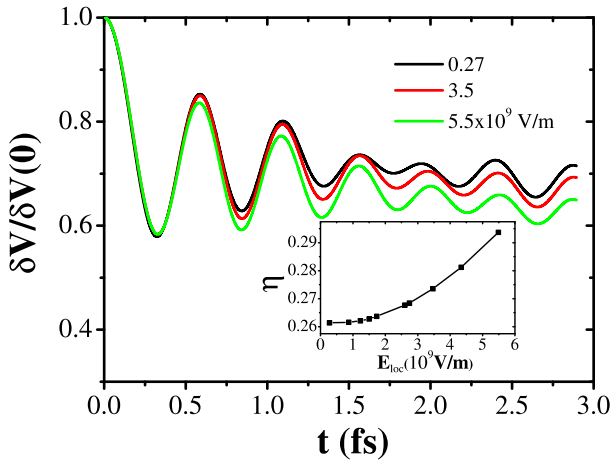}
  \caption{(color online) Time evolutions of the potential difference $\delta V$
  between the silver plates for different enhanced local electric fields
  $E_{loc}$. $d=0.6$ nm; $D=6$ nm. The inset shows the decay rates $\eta$ as a function
  of the local electric field $E_{loc}$. $d=0.6$ nm; $D=6$ nm.}
\label{4}
\end{figure}

It is evident that both the enhancement factor $M$ and the laser
power $P$ can affect the QT via the enhanced local field $E_{loc}$,
which is proportional to $M\sqrt{P}$. We find through numerical
calculations that for the range of laser power commonly used in
experiment, the deciding factor is $M\sqrt{P}$, not individual
values of $M$ and $P$.  For example, we find that the time evolution
of $\delta V$ for $P=10$ mW, $M=100$ and $P=100$ $\mu$W, $M=1000$ is
almost the same (not shown). This means that  we need to consider
only the enhanced local field $E_{loc}$. Figure \ref{4} shows the
time evolutions of $\delta V$ for different $E_{loc}$ at $d=0.6$ nm
and $D=6$ nm. We see clearly that larger local field induce larger
QT, which in turn reduces the enhancement. As shown in the inset in
Fig. \ref{4},  the decay rate $\eta$ decreases slowly when
$E_{loc}<2\times 10^9$ V/m,  and reaches a non-zero constant when
$E_{loc}$ goes to 0. This can be explained by the fact that when the
tunneling is small, we  still have the  linear current-voltage
relation \cite{simmons}, $J(t)=\beta\delta V(t)$. From this
relation, we obtain
\begin{eqnarray}
\delta V(t)=\delta V(0)e^{-d\beta t}\,.
\end{eqnarray}
Therefore, when $\delta V(0)\rightarrow 0$, we have the minimal
decay rate $\eta=1-\delta V(T/2)/\delta V(0)\rightarrow 1-e^{-d\beta
T/2}$. It should be noted that the minimum decay rate  is determined
by the separation. At $d= 0.6$ nm, the minimum reduction is about
$26\%$.


We emphasize that the reduced SERS calculated by us is not
necessarily the overall SERS of a molecule placed in the nano-gap. 
With a molecule in the gap, the situation can become much more complex.    
On the one hand, the oscillatory tunneling current can be coupled
to the molecule inelastically, generating additional Raman signals  \cite{Ho,Johansson}. 
On the other hand, the chemical enhancement can also be affected by the QT \cite{zhao}. 
Thus  the reduced EM enhancement  might be  compensated or even over-
compensated  by these two factors. 
More studies are needed to clarify the issue. 

In sum, we have investigated the time evolution of QT between
two plates using the TDDFT method. We have found that smaller
separation and larger local field  result in stronger QT. Our numrical
results show that when the separation is smaller than $0.6$nm,  the
suppression of  the EM part of enhancement in SERS is very significant
for the common laser power used in experiment.

This work  was supported by NSF of China (10625418, 10825417), by MOST (2005CB724500, 2006CB921400, 2006DFB02020, 
2007CB936800, 2009CB930704), and by the ``Bairen" projects of CAS.


\begin{thebibliography}{21}
\expandafter\ifx\csname natexlab\endcsname\relax\def\natexlab#1{#1}\fi
\expandafter\ifx\csname bibnamefont\endcsname\relax
  \def\bibnamefont#1{#1}\fi
\expandafter\ifx\csname bibfnamefont\endcsname\relax
  \def\bibfnamefont#1{#1}\fi
\expandafter\ifx\csname citenamefont\endcsname\relax
  \def\citenamefont#1{#1}\fi
\expandafter\ifx\csname url\endcsname\relax
  \def\url#1{\texttt{#1}}\fi
\expandafter\ifx\csname urlprefix\endcsname\relax\def\urlprefix{URL }\fi
\providecommand{\bibinfo}[2]{#2}
\providecommand{\eprint}[2][]{\url{#2}}

\bibitem[{\citenamefont{Xu1}(1999)}]{Xu1}
\bibinfo{author}{\bibfnamefont{H. ~X.}~\bibnamefont{Xu}},
\bibinfo{author}{\bibfnamefont{E. ~J.}~\bibnamefont{Bjerneld}},
\bibinfo{author}{\bibfnamefont{M.}~\bibnamefont{K\"{a}ll}},
\bibinfo{author}{\bibfnamefont{L.}~\bibnamefont{B\"{o}rjesson}},
\bibinfo{journal}{Phys. Rev. Lett.} \textbf{\bibinfo{volume}{83}},
\bibinfo{pages}{4357} (\bibinfo{year}{1999}).


\bibitem[{\citenamefont{Mich}(2000)}]{Mich}
\bibinfo{author}{\bibfnamefont{A. ~M.}~\bibnamefont{Michaels}},
\bibinfo{author}{\bibfnamefont{J.}~\bibnamefont{Jiang}},
\bibinfo{author}{\bibfnamefont{L.}~\bibnamefont{Brus}},
\bibinfo{journal}{J. Phys. Chem. B} \textbf{\bibinfo{volume}{104}},
\bibinfo{pages}{11965} (\bibinfo{year}{2000}).


\bibitem[{\citenamefont{Muhlschlegel2005}(2005)}]{Muhlschlegel2005}
\bibinfo{author}{\bibfnamefont{P.}~\bibnamefont{M\"{u}hlschlegel}},
\bibinfo{author}{\bibfnamefont{H. ~J.}~\bibnamefont{Eisler}},
\bibinfo{author}{\bibfnamefont{O. ~J. ~F.}~\bibnamefont{Martin}},
\bibinfo{author}{\bibfnamefont{B.}~\bibnamefont{Hecht}},
\bibinfo{author}{\bibfnamefont{D. ~W.}~\bibnamefont{Pohl}},
\bibinfo{journal}{Science} \textbf{\bibinfo{volume}{308}},
\bibinfo{pages}{1607} (\bibinfo{year}{2005}).

\bibitem[{\citenamefont{Kim2008}(2008)}]{Kim2008}
\bibinfo{author}{\bibfnamefont{S.}~\bibnamefont{Kim}},
\bibinfo{author}{\bibfnamefont{J.}~\bibnamefont{Jin}},
\bibinfo{author}{\bibfnamefont{Y. ~J.}~\bibnamefont{Kim}},
\bibinfo{author}{\bibfnamefont{I. ~Y.}~\bibnamefont{Park}},
\bibinfo{author}{\bibfnamefont{Y.}~\bibnamefont{Kim}},
\bibinfo{author}{\bibfnamefont{S. ~W.}~\bibnamefont{Kim}},
\bibinfo{journal}{Nature} \textbf{\bibinfo{volume}{453}},
\bibinfo{pages}{757} (\bibinfo{year}{2008}).


\bibitem{Raether}      H. H. Raether, Surface Plasmons (Springer, Berlin, 1988).
\bibitem{Mosk}         M. Moskovits, in Top. Appl. Phys. (Springer-Verlag Berlin, Berlin, 2006), pp. 1.


\bibitem[{\citenamefont{Xu2000}(2000)}]{Xu2000}
\bibinfo{author}{\bibfnamefont{H. ~X.}~\bibnamefont{Xu}},
\bibinfo{author}{\bibfnamefont{J.}~\bibnamefont{Aizpurua}},
\bibinfo{author}{\bibfnamefont{M.}~\bibnamefont{K\"{a}ll}},
\bibinfo{author}{\bibfnamefont{P.}~\bibnamefont{Apell}},
\bibinfo{journal}{Phys. Rev. E} \textbf{\bibinfo{volume}{62}},
\bibinfo{pages}{4318} (\bibinfo{year}{2000}).


\bibitem[{\citenamefont{Talley}(2005)}]{Talley}
\bibinfo{author}{\bibfnamefont{ C. ~E.}~\bibnamefont{Talley}},
\bibinfo{author}{\bibfnamefont{J. ~B.}~\bibnamefont{Jackson}},
\bibinfo{author}{\bibfnamefont{C.}~\bibnamefont{Oubre}},
\bibinfo{author}{\bibfnamefont{N. ~K.}~\bibnamefont{Grady}},
\bibinfo{author}{\bibfnamefont{C. ~W.}~\bibnamefont{Hollars}},
\bibinfo{author}{\bibfnamefont{S. ~M.}~\bibnamefont{Lane}},
\bibinfo{author}{\bibfnamefont{T. ~R.}~\bibnamefont{Huser}},
\bibinfo{author}{\bibfnamefont{P.}~\bibnamefont{Nordlander}},
\bibinfo{author}{\bibfnamefont{N. ~J.}~\bibnamefont{Halas}},
\bibinfo{journal}{Nano Lett.} \textbf{\bibinfo{volume}{5}},
\bibinfo{pages}{1569} (\bibinfo{year}{2005}).


\bibitem[{\citenamefont{HWei}(2008)}]{HWei}
\bibinfo{author}{\bibfnamefont{H.}~\bibnamefont{Wei}},
\bibinfo{author}{\bibfnamefont{F.}~\bibnamefont{Hao}},
\bibinfo{author}{\bibfnamefont{Y. ~Z.}~\bibnamefont{Huang}},
\bibinfo{author}{\bibfnamefont{W. ~Z.}~\bibnamefont{Wang}},
\bibinfo{author}{\bibfnamefont{P.}~\bibnamefont{Nordlander}},
\bibinfo{author}{\bibfnamefont{H. ~X.}~\bibnamefont{Xu}},
\bibinfo{journal}{Nano Lett.} \textbf{\bibinfo{volume}{8}},
\bibinfo{pages}{2497} (\bibinfo{year}{2008}).


\bibitem[{\citenamefont{Haes1}(2004)}]{Haes1}
\bibinfo{author}{\bibfnamefont{ A. ~J.}~\bibnamefont{Haes}},
\bibinfo{author}{\bibfnamefont{S. ~L.}~\bibnamefont{Zou}},
\bibinfo{author}{\bibfnamefont{G. ~C.}~\bibnamefont{Schatz}},
\bibinfo{author}{\bibfnamefont{R. ~P.}~\bibnamefont{Van Duyne}},
\bibinfo{journal}{ J. Phys. Chem. B} \textbf{\bibinfo{volume}{108}},
\bibinfo{pages}{109} (\bibinfo{year}{2004}).


\bibitem[{\citenamefont{Haes2}(2004)}]{Haes2}
\bibinfo{author}{\bibfnamefont{ A. ~J.}~\bibnamefont{Haes}},
\bibinfo{author}{\bibfnamefont{S. ~L.}~\bibnamefont{Zou}},
\bibinfo{author}{\bibfnamefont{G. ~C.}~\bibnamefont{Schatz}},
\bibinfo{author}{\bibfnamefont{R. ~P.}~\bibnamefont{Van Duyne}},
\bibinfo{journal}{ J. Phys. Chem. B} \textbf{\bibinfo{volume}{108}},
\bibinfo{pages}{6961} (\bibinfo{year}{2004}).


\bibitem[{\citenamefont{KZhao}(2006)}]{KZhao}
\bibinfo{author}{\bibfnamefont{K.}~\bibnamefont{Zhao}},
\bibinfo{author}{\bibfnamefont{H. ~X.}~\bibnamefont{Xu}},
\bibinfo{author}{\bibfnamefont{B. ~H.}~\bibnamefont{Gu}},
\bibinfo{author}{\bibfnamefont{Z. ~Y.}~\bibnamefont{Zhang}},
\bibinfo{journal}{J. Chem. Phys.} \textbf{\bibinfo{volume}{125}},
\bibinfo{pages}{081102} (\bibinfo{year}{2006}).


\bibitem[{\citenamefont{Xu2007}(2007)}]{Xu2007}
\bibinfo{author}{\bibfnamefont{Z. ~P.}~\bibnamefont{Li}},
\bibinfo{author}{\bibfnamefont{H. ~X.}~\bibnamefont{Xu}},
\bibinfo{journal}{ J. Quant. Spectrosc. Radiat. Transfer} \textbf{\bibinfo{volume}{103}},
\bibinfo{pages}{394} (\bibinfo{year}{2007}).


\bibitem[{\citenamefont{Flatau}(1993)}]{Flatau}
\bibinfo{author}{\bibfnamefont{P. ~J.}~\bibnamefont{Flatau}},
\bibinfo{author}{\bibfnamefont{K. ~A.}~\bibnamefont{Fuller}},
\bibinfo{author}{\bibfnamefont{D. ~W.}~\bibnamefont{Mackowski}},
\bibinfo{journal}{Appl. Opt.} \textbf{\bibinfo{volume}{32}},
\bibinfo{pages}{3302} (\bibinfo{year}{1993}).

\bibitem[{\citenamefont{Futamata}(2003)}]{Futamata}
\bibinfo{author}{\bibfnamefont{M.}~\bibnamefont{Futamata}},
\bibinfo{author}{\bibfnamefont{Y.}~\bibnamefont{Maruyama}},
\bibinfo{author}{\bibfnamefont{M.}~\bibnamefont{Ishikawa}},
\bibinfo{journal}{J. Phys. Chem. B} \textbf{\bibinfo{volume}{107}},
\bibinfo{pages}{7607} (\bibinfo{year}{2003}).


\bibitem[{\citenamefont{Otto}(2002)}]{Otto}
\bibinfo{author}{\bibfnamefont{A.}~\bibnamefont{Otto}},
\bibinfo{journal}{J. Raman Spectrosc.} \textbf{\bibinfo{volume}{33}},
\bibinfo{pages}{593} (\bibinfo{year}{2002}).



\bibitem[{\citenamefont{Danck}(2007)}]{Danck}
\bibinfo{author}{\bibfnamefont{M.}~\bibnamefont{Danckwerts}},
\bibinfo{author}{\bibfnamefont{L.}~\bibnamefont{Novotny}},
\bibinfo{journal}{Phys. Rev. Lett.} \textbf{\bibinfo{volume}{98}},
\bibinfo{pages}{4} (\bibinfo{year}{2007}).

\bibitem[{\citenamefont{Nordlander}(2009)}]{Nordlander}
\bibinfo{author}{\bibfnamefont{J.}~\bibnamefont{Zuloaga}},
\bibinfo{author}{\bibfnamefont{E.}~\bibnamefont{Prodan}},
\bibinfo{author}{\bibfnamefont{P.}~\bibnamefont{Nordlander}},
\bibinfo{journal}{Nano. Lett.} \textbf{\bibinfo{volume}{9}},
\bibinfo{pages}{887} (\bibinfo{year}{2009}).


\bibitem[{\citenamefont{RG}(1984)}]{RG}
\bibinfo{author}{\bibfnamefont{E.}~\bibnamefont{Runge}},
\bibinfo{author}{\bibfnamefont{E. ~K. ~U.}~\bibnamefont{Gross}},
\bibinfo{journal}{Phys. Rev. Lett.} \textbf{\bibinfo{volume}{52}},
\bibinfo{pages}{997} (\bibinfo{year}{1984}).


\bibitem[{\citenamefont{simmons}(1963)}]{simmons}
\bibinfo{author}{\bibfnamefont{J. ~G.}~\bibnamefont{Simmons}},
\bibinfo{journal}{ J. Appl. Phys.} \textbf{\bibinfo{volume}{34}},
\bibinfo{pages}{2581} (\bibinfo{year}{1963}).


\bibitem{cranknicholson}  R. S. Varga, Matrix Iterative Analysis (Prentice-Hall, Englewood Cliffs, NJ, 1962), p. 263.



\bibitem[{\citenamefont{FK}(1976)}]{FK}
\bibinfo{author}{\bibfnamefont{ F. ~K.}~\bibnamefont{Schulte}},
\bibinfo{journal}{Surf. Sci.} \textbf{\bibinfo{volume}{55}},
\bibinfo{pages}{427} (\bibinfo{year}{1976}).

\bibitem{Ho}B. C. Stipe, M. A. Rezaei, W. Ho, Science {\bf 280}, 1732 (1998).

\bibitem{Johansson}P. Johansson, Phys. Rev. B {\bf 58}, 10823 (1998).

\bibitem{zhao} K. Zhao, M. C. Troparevsky, Di Xiao, A. G. Eguiluz, and Zhenyu Zhang, preprint (accepted by PRL)

\end{thebibliography}
\end{document}